\documentclass[aps]{revtex4}
\usepackage{color}
\usepackage{psfrag}
\usepackage{dcolumn}
\usepackage{bm}
\usepackage{float}
\usepackage[latin1]{inputenc}
\usepackage[spanish,english]{babel}
\usepackage{amsfonts}
\usepackage{amssymb,epsf}
\usepackage{graphicx}
\usepackage{epstopdf}
\usepackage{amsmath,amssymb}
\usepackage{hyperref}
\usepackage{appendix}
\begin{document}
\title{The effect of three-body nucleon-nucleon interaction on the ground state binding energy of the light nuclei }
\author{F. Kamgar, G. H. Bordbar\footnote{email address: ghbordbar@shirazu.ac.ir}, S. M. Zebarjad\footnote{email address: zebarjad@shirazu.ac.ir},  M. A. Rastkhadiv}
\affiliation{ Physics Department and Biruni Observatory, Shiraz University, Shiraz 71454, Iran
	}
%
\begin{abstract}
We calculate the ground state binding energies  of the light nuclei such as $^{4}He$, $^{6}Li$, $^{12}C$ and $^{14}N$ by considering the effect of three-body nucleon-nucleon interaction. We use the effective two-body potential obtained from the lowest order constrained variational (LOCV) calculations of  the nuclear matter for the $ Reid68 $, $AV_{14}  $, $UV_{14}$, and $AV_{18} $ nuclear potentials in different channels. To calculate the ground state binding energy, we implement the local density approximation by using the harmonic oscillator wave functions while the effect of  three-body interaction is considered by employing the UIX potential.  We compare the obtained two-body ground state binding energy with the energy related to the three-body effect. We also compare the obtained values with the experimental data and also work of others, and show that the results are relatively acceptable. We   compute the root mean-square radius $R_{rms}$ of the above  nuclei for the $ Reid68 $, $AV_{14}  $, $UV_{14}$, and $AV_{18} $ potentials and compare the results with the experiment. We  also obtain the contribution of different channels by matching to the experimental values of the quadrupole moments and magnetic dipole moments.  Furthermore, we  calculate the three-body cluster energy of the above nuclei and compare the results with that of nuclear matter. According to the obtained results, we see that the three-body cluster energy contribution is small. For example, for $^{4}He$ nuclide, this value is 0.079 MeV with the $ Reid68 $ potential.
\end{abstract}



\maketitle

\section{\large{Introduction}}
\label{Intro}
The three-body potential has a long history in physics, including tidal interaction resulting from the effective three-body interaction potential. In quantum mechanics, a three-body interaction is created by removing degrees of freedom. For  example, a three-body potential obtained from  the two-pion exchange interaction between nucleons has been studied by Fujita-Miyazawa \cite{b1}. Phenomenological studies show that, for example, the contribution of three-body nuclear interaction to the energy of light nuclei is quantitatively about 20 percent \cite{b2}. One of the phenomenological three-nucleon potentials is the Urbana-Argonne (UA) model which consists of a two-pion-exchange term and a repulsive term \cite{b3,b4}.

 Pandharipande proposed the lowest order constrained variational (LOCV) method in 1971  to solve the  problems consisting  of fermions (nuclear and neutron matter and $ ^{3}He $) and bosons ($ ^{4}He $) \cite{b5,b6,b7}. In this method, two-body correlation functions were extracted from the particle energy minimization in the approximation of two-body clusters in the presence of a constraint. Pandharipande's method was not entirely a variational method \cite{b8,b9} and therefore, Irvine et al. \cite{b10} modified the LOCV method to calculate the energy of a neutron matter (liquid $ ^{3}He $). The presence of intermediate states in nucleon-nucleon interactions for light nuclei  was investigated in Refs. \cite{b11, b14}.
 The magnetic susceptibility of neutron matter was calculated \cite{b13}, and then it was generalized to finite temperature \cite{b15}.  In 1997, the properties of Asymmetric nuclear matter was investigated at finite temperatures \cite{b16}. The LOCV method also was formulated for $ UV_{14} $ interaction \cite{b17} agrees  with the results  corresponding to FHNC calculations \cite{b18}. In 1998, the LOCV method was formulated for the two-body modern
  $ AV_{18} $ potential, and asymmetric nuclear matter calculations were reported in Ref. \cite{b19}. The properties of asymmetric nuclear matter at finite temperatures were then calculated for $ AV_{18} $ potential \cite{b20}.
  Furthermore the  magnetized hot neutron matter  \cite{b22},  Thermodynamic and saturation properties for hot magnetized nuclear matter \cite{b23}, and spin polarized asymmetric nuclear matter and neutron star matter \cite{b24,b25} were obtained by this method. The Lowest order constrained variational and local density approximation approach to the hot alpha particle was considered in Ref.\cite{b60}. In 2005,  the ground state of light and heavy closed-shell nuclei was also investigated by the method of effective interaction and local density approximation in Refs.\cite{b70,b71}. Also in this year,  the angular momentum dependent calculation and  the effect of spin-dependent correlation functions on the ground state energy of liquid $^{3}He$ was studied in Refs.   \cite{b61,b62}. In 2011, the effect of density-dependent AV18 effective interactions on the ground state properties of heavy closed shell nuclei was calculated \cite{b72}. Recently, the effects of three-body interaction on the nuclear matter properties and neutron star structure have been calculated \cite{b26,b26-2}.

  The LOCV method has been also used for light nuclei \cite{b12}. There was some difficulty defining the correlation function's long-range behavior, and the computational results were not satisfactory for closed and light shell nuclei. To overcome this difficulty, the local density approximation was used to create an effective two-body potential for LOCV calculations  related to the nuclear matter at different densities \cite{b27}.
   In the present work, by employing the three-body nuclear potential,  we calculate the ground state binding energy for the nuclei $^{4}He$, $^{6}Li$, $^{12}C$ and $^{14}N$ by using the harmonic oscillator bases and the local density approximation. Furthermore, we compute the three-body cluster energy for the mentioned  nuclei.  In our calculations, we use the $ Reid68 $, $AV_{14}  $, $UV_{14}$, and $AV_{18} $ potentials. The $ Reid68 $ potential contains three terms, central, tensor, and spin-orbital, and was first introduced in 1968 by Reid \cite{b200}. The $UV_{14}$ potential, introduced by Lagaris and Pandharipande in 1981 \cite{b201}, is the fully phenomenological potential that includes 14 operators. Wiringa et. al. proposed the  $AV_{14} $ potential \cite{b202}, which  also contains a 14-operator potential that  differs from UV14 in the definition of the functions in the potential. The $AV_{18} $ \cite{b2} is an exceptional potential which incorporates the charge dependent  and charge asymmetry. This model includes an independent part with 14 operators, which is an updated version of the $AV_{14} $ potential. Furthermore, three additional charge-dependent operators and a charge asymmetry operator with a full electromagnetic interaction are added to the $AV_{18} $ potential. The $AV_{18} $ potential is consistent with pp and nn data as well as low-energy nn parameters and deuteron properties.  Using the above potentials, in Ref. \cite{b19},  we computed the  nuclear matter saturation properties. The energy results are given in Table\ref{j5}. According to the results, it is seen that the  most attractive  potential belongs to   $Reid68$ potential and the less attractive potential is related to   $AV_{14}$ potential. Here we also see that by adding the effect of three-nucleon interaction ($ UV14+TNI$), the obtained saturation energy is in good agreement with that  of experimental data.
  \begin{table}[h!]
 	\begin{center}	
 			\caption{The saturation energy of nuclear matter for different nuclear potentials  \cite{b19}.}
 			\label{j5}
 			
 			\begin{tabular}{|c|c|c|c|}
 				\hline
 				
 				$ Potential $&$ E(\rho_{0})(MeV)$	 \\
 				\hline
 				$ Reid $&$ -22.83$\\
 					$ UV14 $&$ -21.20$\\
 				$ AV18 $&$ -18.46$\\
 				
 				$ AV14 $&$ -15.99$\\
 					\hline 				
 				$ UV14+TNI $&$ -17.33$\\
 					\hline 	
 			 Experiment &$ -16\pm 1 $\\
 				
 				\hline
 				
 			\end{tabular}
 	\end{center}
 \end{table}

The structure of this paper is as follows: First, we present our method for calculating the ground state binding energy of light nuclei in section \ref{method}. In section \ref{three}, we explain our technique for computing the effect of three-body nuclear potential by using the UIX model \cite{b4}. The results are presented in  sections \ref{results}.  In section \ref{E3}, the three-body cluster energy value is calculated. The summary and discussion are given in Section \ref{summary}.

\section{\large{ The light nuclei ground state binding energy  }}
\label{method}
 Here, we use a variational method to calculate the energy of system up to the effect of two-body interaction. The Hamiltonian for a system of $N$ particles up to the two-body potential is as follows:
\begin{eqnarray}
H_{2}=\sum_{i=1}^ {N} -\frac{\hbar^2}{2m}\triangledown_{i}^2+\sum_{i<j}V_{ij},
\label{eq1}
\end{eqnarray}
 where $ V_{ij} $ is the two-body potential. The main problem of the variational method is to calculate the Hamiltonian expectation value using an trial wave function $ \Psi=F\Phi $, where $ \Phi $ is the wave function of the system with $N$ non-interacting particles and $ F $ is the correlation function. $ F $ is the correction due to the inter particle interaction and provides a correlation between particles. The energy expectation value is:
\begin{eqnarray}
E=\frac{\langle \psi |H_{2}|\psi  \rangle}{\langle \psi | \psi \rangle},
\end{eqnarray}
It is impossible to solve the above relation for large $ N$, so here we use a cluster expansion for the energy expectation value which up to the two-body term is as follows \cite{b32},
\begin{equation}
E=E_1 + E_2,
\label{energy}
\end{equation}
where $E_1$ is the one-body energy and $E_2$ is the two-body cluster energy,
\begin{equation}
	E_{2}=\frac{1}{2N}\sum_{i,j}\langle {ij}| {v(12)}|{ij-ji}\rangle.
\end{equation}
In above equation, $v(12)$ is the two-body effective potential,
\begin{equation}
	 {v(12)}=-\frac{\hbar ^{2}}{2m}\big[f(12),\big[\bigtriangledown
	_{12}^{2},f(12)\big]\big]+f(12)V(12)f(12),
	\label{eq5}
\end{equation}
where $f(12)$ and $V (12)$ are the two-body correlation function and two-body interaction potential, respectively.
 By performing a series of mathematical calculations, the two-body energy can be calculated. By minimizing the energy relationship concerning the correlation function and using the Euler-Lagrange equations, we arrive the second-order differential equation for the correlation function.
 %
%
{For solving the mentioned differential equation, we impose the following boundary condition on the correlation function \cite{b17},
\begin{equation}
0\leq f(r) \leq f_p(k_Fr),
\end{equation}
where $ f_p(k_F r) = \{1 - \frac{9}{4}[\frac{J_1(k_F r}{k_F r}]^2\}^{-1/2}$ is the Pauli function.
Here, $J_{1}(x)$ is the spherical Bessel function, and the Fermi momenta $ k_{F} $ is fixed by the density of system, $ k_{F}=(\frac{6\pi^{2}}{4}\rho)^{\frac{1}{3}} $}.
By using the obtained correlation function in the above equations, it is possible to obtain the two-body cluster energy.

According to the calculations in the LOCV method, we obtain an operator form for the effective potential in each channel as follows:
\begin{eqnarray}\label{poteff}
v_{\alpha}(r,k_{f})=-\frac{\hbar^{2}}{2m}\big[f_{\alpha}(r),\big[\triangledown^{2}_{r},f_{\alpha}(r)\big]\big]+f_{\alpha}(r)V_{\alpha}(r)f_{\alpha}(r),
\end{eqnarray}
where  $ f_{\alpha}(r) $ is the correlation function in channel $\alpha$ and  $ k_{f} $ is the fermi momentum of nucleons.  $ V_{\alpha}(r) $ is  a real two-body potential considered to be  $ Reid68 $, $AV_{14}  $, $UV_{14} $, and $AV18$.

To calculate the ground state binding energy, we consider the single particle wave function as a harmonic oscillator wave functions ($\alpha\equiv n,l$) to determine the matrix element for the effective two-body potential:
\begin{eqnarray}
\label{2}
m_{\alpha}(\hbar \omega,k_{f})=\langle nl|v_{\alpha}(r,k_{f})|nl\rangle,
\end{eqnarray}
where $ \hbar \omega$  is the harmonic oscillator parameter, and $ r=|\textbf{r}_{1}-\textbf{r}_{2}| $ is the relative distance. Eq. (\ref{2}) can be calculated in different channels for different nuclei.

We  then use the local density approximation to calculate the interaction energy in each channel and the ground state binding energy by using the following assumption:
\begin{eqnarray}
\frac{1}{2}\big[\rho (\textbf{r}_{1})+\rho (\textbf{r}_{2})\big]\simeq\rho \big(\frac{\textbf{r}_{1}+\textbf{r}_{2}}{2}\big).
\end{eqnarray}
In the above equation,  we approximate the average density of nucleons at $ \textbf{r}_{1} $ and $ \textbf{r}_{2} $ points with their center of mass density. Moreover, the relationship between density and the harmonic oscillator wave function is known to be:
\begin{eqnarray}
\rho\big(\frac{\textbf{r}_{1}+\textbf{r}_{2}}{2}\big)=\big(\frac{2}{3\pi^{2}}\big)k_{f}^{3}=\big|\phi_{nl}\big(\frac{R}{\sqrt{2}}\big)\big|^{2},
\end{eqnarray}
where
\begin{eqnarray}
\textbf{R}=\frac{\textbf{r}_{1}+\textbf{r}_{2}}{\sqrt{2}}.
\end{eqnarray}

Now, we compute the ground state binding energy per nucleon up to the two-body  potential contribution,
\begin{eqnarray}
BE_2=\frac{1}{N}(T+E_{\alpha}), \label{6}
\end{eqnarray}
where $ E_{\alpha} $ is obtained by summing over the spin and isospin states in the $ \alpha $ channel:
\begin{eqnarray}
E_{\alpha}=\sum_{m_{s},m_{\tau}}e_{\alpha},
\end{eqnarray}
and $e_{\alpha}$ is the interaction energy in each channel calculated by integrating  the center of mass  coordinates:
\begin{eqnarray}
e_{\alpha}=\int_{0}^{\infty}\big|\phi_{nl}(\frac{R}{\sqrt{2}})\big|^{2}m_{\alpha}(\hbar \omega,k_{f})R^{2}dR.
\end{eqnarray}
In Eq. (\ref{6}), the kinetic term $ T $ is obtained by:
\begin{eqnarray}
T=\frac{1}{2}\sum_{i=1}^{N}\big(2n_{i}+l_{i}+\frac{3}{2}\big)\hbar \omega - T_{c.m.},
\end{eqnarray}
where $T_{c.m.}=\frac{3}{4}\hbar \omega$ is center of mass energy.
For example the ground state binding energy per nucleon for the nuclide $^{4}He$ is given by:
\begin{eqnarray}
BE_2=\frac{1}{N}\left (3\hbar \omega -\frac{3}{4}\hbar \omega +3e_{(^{3}S_{1})}+3e_{(^{1}S_{0})}\right ).
\end{eqnarray}


\section{\large{ Ground state binding energy by considering  the three-body potential contribution }}
\label{three}

In previous section, the ground state binding energy calculations for the light nuclei (such as $^{4}He$, $^{6}Li$, $^{12}C$ and $^{14}N$) was specified by employing two-body nuclear potential.
In this section our aim is to implement the effect of three-body nucleon-nucleon potential in the above calculations.
The Hamiltonian for calculations is as follows:
\begin{eqnarray}
H=H_{2}+\sum\limits_{i<j<k\leq N}V_{ijk},
\end{eqnarray}
where the first term has been given in the previous section (Eq. (\ref{energy})) and $ V_{ijk} $ is the three-body potential.
Considering the effect of three-body potential, the energy of system can be computed as follows,
\begin{equation}\label{v31}
E=E_{1}+E_2+\langle\sum\limits_{ijk}V_{ijk}\rangle.
\end{equation}
The computation of the first two terms has been discussed in previous section. Here, we  calculate the contribution of three-body potential,
\begin{eqnarray}&&
\langle V_{3}\rangle=\langle\sum\limits_{ijk}V_{ijk}\rangle=\frac{\langle \psi |\sum_{ijk}V_{ijk}|\psi  \rangle}{\langle \psi | \psi \rangle}\nonumber\\&&
=\sum\limits_{\sigma_{1}...\sigma_{N}}\sum\limits_{\tau_{1}...\tau_{N}}\int...\int d\mathbf{r}_{1}...d\mathbf{r}_{N}\psi^{\ast}(\mathbf{r}_{1}\sigma_{1}\tau_{1},...,\mathbf{r}_{N}\sigma_{N}\tau_{N})
\sum\limits_{i<j<k}V(\mathbf{r}_{i},\mathbf{r}_{j},\mathbf{r}_{k})\psi(\mathbf{r}_{1}\sigma_{1}\tau_{1},...,\mathbf{r}_{N}\sigma_{N}\tau_{N})\nonumber\\&&
=\frac{N(N-1)(N-2)}{3!}\sum\limits_{\sigma_{1}...\sigma_{N}}\sum\limits_{\tau_{1}...\tau_{N}}\int...\int d\mathbf{r}_{1}...d\mathbf{r}_{N}
\psi^{\ast}(\mathbf{r}_{1}\sigma_{1}\tau_{1},...,\mathbf{r}_{N}\sigma_{N}\tau_{N}) V(\mathbf{r}_{1},\mathbf{r}_{2},\mathbf{r}_{3})\nonumber\\&& \times\psi(\mathbf{r}_{1}\sigma_{1}\tau_{1},...,\mathbf{r}_{N}\sigma_{N}\tau_{N}).
\label{v3}
\end{eqnarray}
In the above equation, $V(\mathbf{r}_{1},\mathbf{r}_{2},\mathbf{r}_{3}) $ is the three-body potential.
For the UIX model, the three-body potential consists of two terms in the form of $ V_{ijk}=V_{ijk}^{2\pi}+V_{ijk}^{R} $. The long-range $ V_{ijk}^{2\pi} $ is an attractive and was first suggested by Fujita and Miyazawa \cite{b1,b18,b28} as:
\begin{eqnarray}&&
V_{ijk}^{2\pi}=A_{2\pi}\sum_{cyc}\bigg(\{X_{ij}^{\pi},X_{ik}^{\pi}\}\{ \mathbf{\tau_{i}}.\mathbf{\tau_{j}},\mathbf{\tau_{i}}.\mathbf{\tau_{k}}\}\nonumber\\&&
\,\,\,\,\,\,\,\,\,\,\,\,\,\,\,\,\,\,\,+\frac{1}{4}[X_{ij}^{\pi},X_{ik}^{\pi}][ \mathbf{\tau_{i}}.\mathbf{\tau_{j}},\mathbf{\tau_{i}}.\mathbf{\tau_{k}}]\bigg),
\end{eqnarray}
where $ X_{ij}^{\pi} $ and $\mathbf{S}_{ij}$ are
\begin{eqnarray}
X_{ij}^{\pi}&=& Y_{\pi}(r_{ij})\mathbf{\sigma}_{i}.\mathbf{\sigma}_{j}+T_{\pi}(r_{ij})\mathbf{S}_{ij},\nonumber\\
\mathbf{S}_{ij}&=&3(\mathbf{\sigma}_{i}.\mathbf{\hat{r}}_{ij})(\mathbf{\sigma}_{j}.\mathbf{\hat{r}}_{ij})-\mathbf{\sigma}_{i}.\mathbf{\sigma}_{j}.
\end{eqnarray}
$ Y_{\pi}(r) $ and $ T_{\pi}(r) $ are the radial functions associated with the Yukawa and tensor parts of one pion exchange interaction:
\begin{eqnarray}
Y_{\pi}(r)&=&\frac{e^{-m_{\pi}r}}{m_{\pi}r}\big(1-e^{-cr^{2}}\big),
\nonumber\\
T_{\pi}(r)&=&\frac{e^{-m_{\pi}r}}{m_{\pi}r}\big(1+\frac{3}{m_{\pi}r}+\frac{3}{m_{\pi}^{2}r^{2}}\big)\big(1-e^{-cr^{2}}\big)^{2},
\end{eqnarray}
where $\mu=0.7fm^{-1} $ is the average mass of the pion, $ c=2.1fm^{-2} $ is the cut-off constant for the   $ V_{14} $ potentials \cite{b2,b30}, and $ A_{2\pi} $ is equal to $ -0.0329 $ and $ -0.0331 $ for the $ AV_{14} $ and $ UV_{14} $ potentials , respectively.
The second part of the three-body potential $ V_{ijk}^{R} $ includes the intermediate-range repulsive part \cite{b201},
\begin{eqnarray}
V_{ijk}^{R}=U\sum_{cyc}T_{\pi}^{2}(r_{ij})T_{\pi}^{2}(r_{ik}),
\end{eqnarray}
where $ U $ is equal to $ 0.0064 $ and $ 0.0045 $ for the $ AV_{14} $ and $ UV_{14} $ potentials , respectively.

Now, by defining a three-body density matrix $ \Gamma(\mathbf{r}_{1}\sigma_{1}\tau_{1},\mathbf{r}_{2}\sigma_{2}\tau_{2},\mathbf{r}_{3}\sigma_{3}\tau_{3}) $ as:
\begin{eqnarray}
\Gamma(\mathbf{r}_{1}\sigma_{1}\tau_{1},\mathbf{r}_{2}\sigma_{2}\tau_{2},\mathbf{r}_{3}\sigma_{3}\tau_{3})=&&  \frac{N(N-1)(N-2)}{3!}
\sum\limits_{\sigma_{4}...\sigma_{N}}\sum\limits_{\tau_{4}...\tau_{N}} \int...\int d\mathbf{r}_{4}...d\mathbf{r}_{N} \nonumber\\&& \times \ \psi^{\ast}(\mathbf{r}_{1}\sigma_{1}\tau_{1},\mathbf{r}_{2}\sigma_{2}\tau_{2},...,\mathbf{r}_{N}\sigma_{N}\tau_{N}) \psi(\mathbf{r}_{1}\sigma_{1}\tau_{1},\mathbf{r}_{2}\sigma_{2}\tau_{2},...,\mathbf{r}_{N}\sigma_{N}\tau_{N}),
\end{eqnarray}
and also a three-body radial distribution function as:
\begin{eqnarray}
\rho^{3}g(\mathbf{r}_{1},\mathbf{r}_{2},\mathbf{r}_{3})=3!\sum\limits_{\sigma_{1}\sigma_{2}\sigma_{3}}
\sum\limits_{\tau_{1}\tau_{2}\tau_{3}}\Gamma(\mathbf{r}_{1}\sigma_{1}\tau_{1},\mathbf{r}_{2}\sigma_{2}\tau_{2},\mathbf{r}_{3}\sigma_{3}\tau_{3}),\nonumber\!\!\!\!\!\!\!\! \label{14}\\
\end{eqnarray}
we can rewrite the expectation value of three-body potential, Eq. (\ref{v3}), as follows:
\begin{eqnarray}
\langle V_{3}\rangle=\frac{\rho^{3}}{6}\int \int \int d\mathbf{r}_{1}d\mathbf{r}_{2}d\mathbf{r}_{3} V(\mathbf{r}_{1},\mathbf{r}_{2},\mathbf{r}_{3})g(\mathbf{r}_{1},\mathbf{r}_{2},\mathbf{r}_{3}).\nonumber\\ \label{15}
\end{eqnarray}
A self-consistent method can be used to obtain the three-body radial distribution function as follow \cite{b32}:
\begin{eqnarray}
g(\mathbf{r}_{1},\mathbf{r}_{2},\mathbf{r}_{3})=f^{2}(r_{12})f^{2}(r_{13})f^{2}(r_{23}) g_{F}(\mathbf{r}_{1},\mathbf{r}_{2},\mathbf{r}_{3}).\label{20}
\end{eqnarray}
In the above equation, $g_{F}(\mathbf{r}_{1},\mathbf{r}_{2},\mathbf{r}_{3})$ is the three-body radial distribution function of the non-interacting system.
In Dirac approach, the total wave function of the system is written by the slater determinant of single particle wave function \cite{b41},
\begin{eqnarray}
\Phi(\mathbf{r}_{1}\sigma_{1}\tau_{1},...,\mathbf{r}_{N}\sigma_{N}\tau_{N})=\frac{1}{\sqrt{N!}}det\big(\phi_{i}(\mathbf{r}_{j}\sigma_{j}\tau_{j})\big),
\end{eqnarray}
where $\phi_{i}$ is the single particle wave function.
Using Eqs. (\ref{14}) and (\ref{20}), the non-interacting three-body radial distribution function is obtained as follows:
\begin{eqnarray}
g_{F}(\mathbf{r}^{\prime}_{1},\mathbf{r}^{\prime}_{2},\mathbf{r}^{\prime}_{3};\mathbf{r}_{1},\mathbf{r}_{2},\mathbf{r}_{3})=\frac{1}{3!}\begin{vmatrix} \gamma(\mathbf{r}^{\prime}_{1},\mathbf{r}_{1}) & \gamma(\mathbf{r}^{\prime}_{1},\mathbf{r}_{2}) & \gamma(\mathbf{r}^{\prime}_{1},\mathbf{r}_{3}) \\ \gamma(\mathbf{r}^{\prime}_{2},\mathbf{r}_{1}) & \gamma(\mathbf{r}^{\prime}_{2},\mathbf{r}_{2}) & \gamma(\mathbf{r}^{\prime}_{2},\mathbf{r}_{3}) \\ \gamma(\mathbf{r}^{\prime}_{3},\mathbf{r}_{1}) & \gamma(\mathbf{r}^{\prime}_{3},\mathbf{r}_{2}) & \gamma(\mathbf{r}^{\prime}_{3},\mathbf{r}_{3}) \end{vmatrix},\nonumber\\
\end{eqnarray}
where $ \gamma(\mathbf{r}^{\prime},\mathbf{r})=\sum\limits_{i} \phi^{*}_{i}(\mathbf{r}^{\prime})\phi_{i}(\mathbf{r})$. Here, the diagonal element, $g_{F}(\mathbf{r}_{1},\mathbf{r}_{2},\mathbf{r}_{3})$  represents the three-body probability density. In the harmonic oscillator approximation $ \phi $ is as follows:
\begin{eqnarray}
\phi_{nlm}(\mathbf{r},\theta,\phi)=R_{nl}(\mathbf{r})Y_{l}^{m}(\theta,\phi).
\end{eqnarray}
After some algebra, we get $g_{F}(\mathbf{r}_{1},\mathbf{r}_{2},\mathbf{r}_{3})$ as:
\begin{eqnarray}
g_{F}(\mathbf{r}_{1},\mathbf{r}_{2},\mathbf{r}_{3})&=&1-\frac{1}{\nu}\big(e^{-\gamma^{2}r^{2}}+e^{-\gamma^{2}r^{\prime2}}
+e^{-\gamma^{2}(r^{2}+r^{\prime2}-2rr^{\prime}cos(\theta))}\big)\nonumber\\&&
+\frac{2}{\nu^{2}}\big(e^{-\gamma^{2}(r^{2}+r^{\prime2}-rr^{\prime}cos(\theta))}\big). \label{24}
\end{eqnarray}
Finally, using Eqs. (\ref{24}), (\ref{20}) and (\ref{15}), we can calculate the ground state binding energy per nucleon including the three-body potential contribution for the nuclei,
\begin{equation}
	BE= BE_2+\frac{1}{N} \langle V_{3}\rangle.
\end{equation}
%

{Here, it should be mentioned that in our present calculations, the contribution of three-nucleon interaction energy is separately computed using the three-body distribution function based on the two-body correlation function which has been obtained based on the two-body interaction. Therefore, our contribution of three-nucleon potential can be treated similar to that of the perturbation theory.}
The results are presented in the next section.

	\section{\large{Results}}
\label{results}
\begin{figure}[h!]
	\begin{center}
		\includegraphics[scale=0.63]{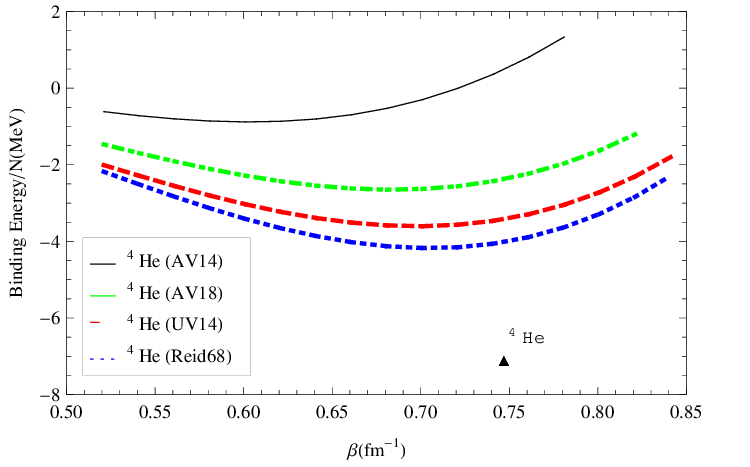}
		\caption{$BE_2$ for the nuclide $^{4}He$  vs $\beta = \sqrt{\frac{m\omega}{\hbar}}$.  The solid line curve corresponds to $^{4}He$ energy with the $ AV_{14} $ potential, the dashed line corresponds with $ UV_{14} $ potential, and the dotted curve corresponds with $Reid68$ potential. Black Up-pointing Triangle is experimental $^{4}He$ prediction \cite{b42}.}
		\label{b1}
	\end{center}
\end{figure}

\begin{figure}[h!]
	\begin{center}
		\includegraphics[scale=0.65]{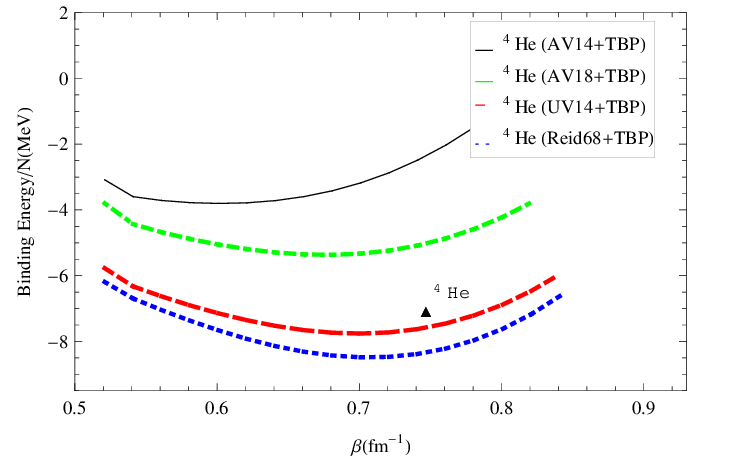}
		\caption{$BE$ for the nuclide $^{4}He$ vs $\beta = \sqrt{\frac{m\omega}{\hbar}} $. The solid line curve corresponds to $^{4}He$ energy with the $ AV_{14} $ potential, the dashed line corresponds with $ UV_{14} $ potential, and the dotted curve corresponds with $Reid68$ potential. Black Up-pointing Triangle is experimental $^{4}He$ prediction \cite{b42}.}
		\label{b2}
	\end{center}
\end{figure}

\begin{figure}[h!]
	\begin{center}
		\includegraphics[scale=0.65]{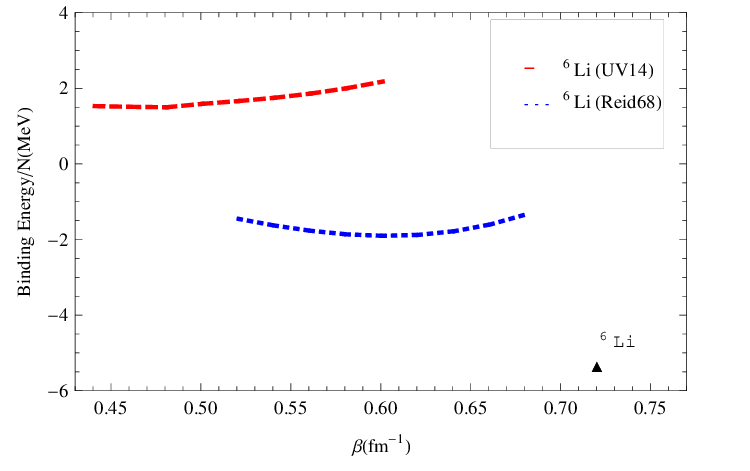}
		\caption{$BE_2$ for the nuclide $^{6}Li$  vs $\beta = \sqrt{\frac{m\omega}{\hbar}} $. The dashed line corresponds with $ UV_{14} $ potential, and the dotted curve corresponds with $Reid68$ potential. Black Up-pointing Triangle is experimental $^{6}Li$ prediction \cite{b42}.}
		\label{b3}
	\end{center}
\end{figure}

\begin{figure}[h!]
	\begin{center}
		\includegraphics[scale=0.65]{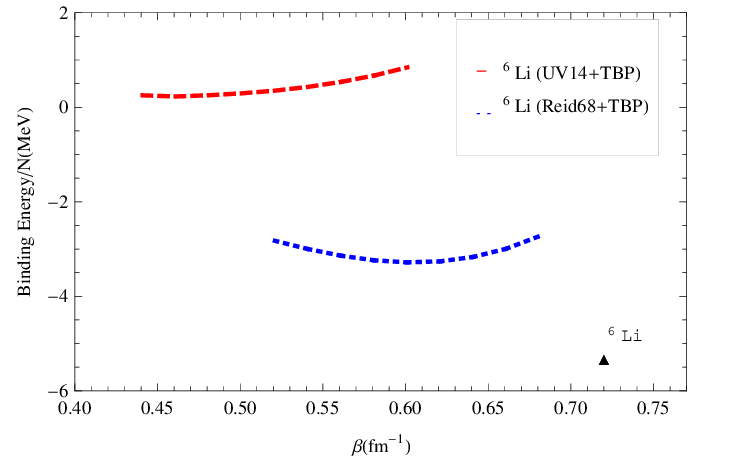}
		\caption{$BE$ for the nuclide $^{6}Li$  vs $\beta = \sqrt{\frac{m\omega}{\hbar}} $. The dashed line corresponds with $ UV_{14} $ potential, and the dotted curve corresponds with $Reid68$ potential.Black Up-pointing Triangle is experimental $^{6}Li$ prediction \cite{b42}.}
		\label{b4}
	\end{center}
\end{figure}

\begin{figure}[h!]
	\begin{center}
		\includegraphics[scale=0.65]{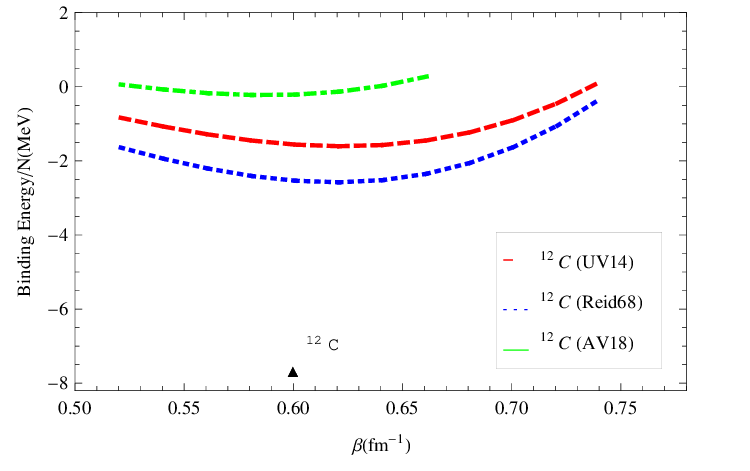}
		\caption{$BE_2$ for the nuclide $^{12}C$ vs $\beta = \sqrt{\frac{m\omega}{\hbar}} $. The dashed line corresponds with $ UV_{14} $ potential, and the dotted curve corresponds with $Reid68$ potential.Black Up-pointing Triangle is experimental $^{12}C$ prediction \cite{b42}.}
		\label{b5}
	\end{center}
\end{figure}

\begin{figure}[h!]
	\begin{center}
		\includegraphics[scale=0.65]{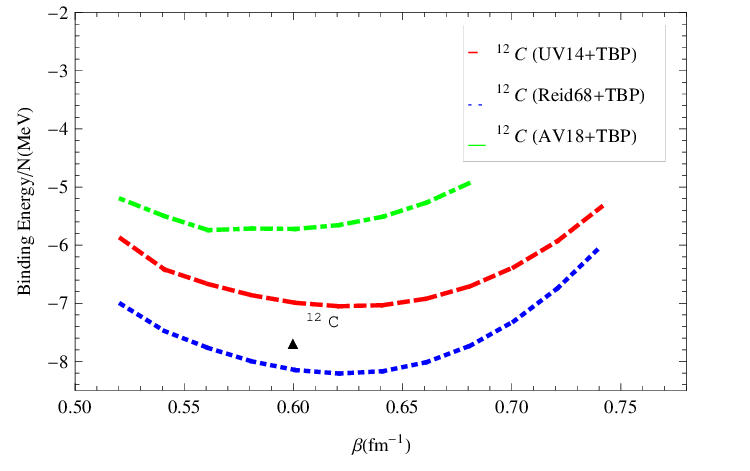}
		\caption{$BE$ for the nuclide $^{12}C$  vs $\beta = \sqrt{\frac{m\omega}{\hbar}} $. The dashed line corresponds with $ UV_{14} $ potential, and the dotted curve corresponds with $Reid68$ potential. Black Up-pointing Triangle is experimental $^{12}C$ prediction \cite{b42}.}
		\label{b6}
	\end{center}
\end{figure}

\begin{figure}[h!]
	\begin{center}
		\includegraphics[scale=0.65]{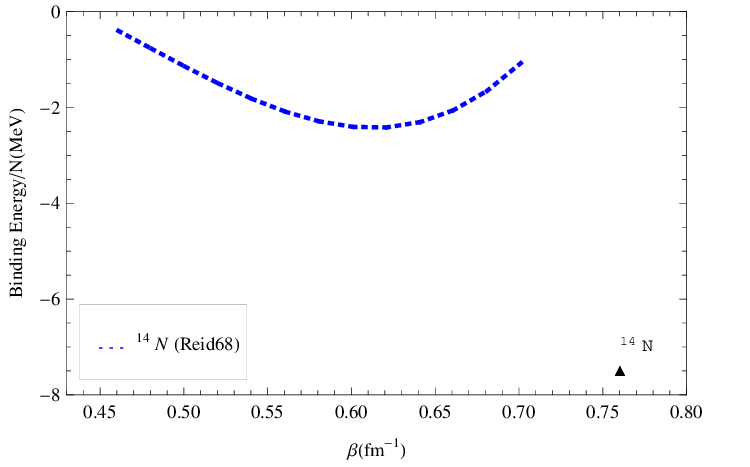}
		\caption{$BE_2$ for the nuclide $^{14}N$ vs $\beta = \sqrt{\frac{m\omega}{\hbar}} $. The dotted curve corresponds with $Reid68$ potential.Black Up-pointing Triangle is experimental $^{14}N$ prediction \cite{b42}.}
		\label{b7}
	\end{center}
\end{figure}

\begin{figure}[h!]
	\begin{center}
		\includegraphics[scale=0.65]{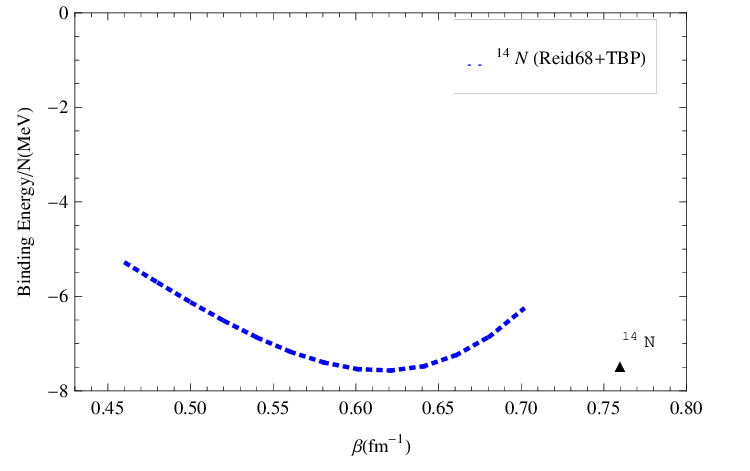}
		\caption{$BE$ for the nuclide $^{14}N$  vs $\beta = \sqrt{\frac{m\omega}{\hbar}} $. The dotted curve corresponds with $Reid68$ potential. Black Up-pointing Triangle is experimental $^{14}N$ prediction \cite{b42}.}
		\label{b8}
	\end{center}
\end{figure}

The results of our calculations for the energy per nucleon of the nuclei $^{4}He$, $^{6}Li$, $^{12}C$ and $^{14}N$ as a function of $\beta = \sqrt{\frac{m\omega}{\hbar}}$ are shown in Figs. \ref{b1}-\ref{b8} for two cases without ($BE_2$) and with ($BE$) the effect of the three-body nuclear potential. For all figures, the energy curves show a minimum at the specific values of $\beta$. For each nuclide, the ground state binding energy per nucleon corresponds to the relevant minimum point.

Table \ref{j1} show the results for the ground state binding energy per nucleon without ($BE_2$) and with ($BE$) the effect of the three-body potential contribution for the above nuclei, respectively. Here a comparison has been also made with the experimental data. By comparing Table \ref{j1} with Table \ref{j5}, we see that  the attractive behavior of employed potentials for finite nuclie is similar to those for nuclear matter. This correspondence is also clearly visible when we include the effect of the three-body force.

 We have also computed the root mean-square radius $R_{rms}$ for these nuclei,
\begin{eqnarray}
R_{rms}=\langle r^{2}\rangle^{\frac{1}{2}}=\dfrac{\int r^{2}|\psi_{nl}(\mathbf{r})|^{2}d\mathbf{r}}{\int |\psi_{nl}(\mathbf{r})|^{2}d\mathbf{r}}.
\label{31}
\end{eqnarray}
Our results are given in Table \ref{j1} and compared with those of experiment \cite{b43}.

We have also compared our results with those of others in Table \ref{j3}. According to these tables, we see that in the case of the nuclide $^{4}He$, the value of energy with the $ UV_{14} $ potential is close to the experimental value and the results of others. By introducing the three-body potential effect, our results approach the experiment values. In the case of the  nuclide $^{6}Li$, the result related to the $ Reid68 $ potential is almost close to results of others. For the nuclide $^{12}C$, our results with both $ Reid68 $ and $ UV_{14} $ potentials are closer to the experimental data {and those of GFMC with the $AV18+IL7$ potential \cite{b222}} with respect to the results of other works.
 \begin{table*}[ht]
 \begin{center}	
 	\begin{scriptsize}
        	\caption{The ground state binding energy per nucleon and the corresponding $\beta = \sqrt{\frac{m\omega}{\hbar}} $ without ($BE_2$) and with ( $BE$) the three-body potential contribution for the nuclei $^{4}He$, $^{6}Li$, $^{12}C$ and $^{14}N$ and the corresponding values of root mean-square radius ($R_{rms}$)  employing different potentials. The corresponding experimental data (Exp.) for the ground state binding energy per nucleon \cite{b42} and $R_{rms}$ \cite{b43} have been also given for comparison.}
			\label{j1}

			\begin{tabular}{|c|c|c|c|c|c|c|c|}
				\hline
				$ Nuclide $&$ Potential $&$\beta(fm^{-1}) $&$ BE_2 \,(MeV)$& $BE\,(MeV)$&Exp.\ $(MeV)$ &$ R_{rms}(fm) $&$ R_{rms}^{Exp}(fm) $	 \\
				\hline
				{$^{4}He$}&$ Reid68 $&$0.701 $&$ -4.173$&$ -8.484$&$-7.08$&$ 1.747 $&$ 1.675 $\\
			
				&$ UV14 $&$0.701 $&$ -3.603$&$ -7.761$&$  $&$ 1.747 $&$  $\\
					&$ AV18 $&$0.681 $&$ -2.65$&$ -5.367$&$  $&$ 1.798 $&$  $\\
				&$ AV14 $&$0.601 $&$ -0.881$&$ -3.803$&$  $&$ 2.037 $&$  $\\
			
				\hline
				{$^{6}Li$}&$ Reid68 $&$0.601 $&$ -1.902$&$ -3.282$&$-5.33$&$ 2.04 $&$ 2.58 $\\
				
				&$ UV14 $&$0.481 $&$ 1.5$&$ 0.229$&$  $&$ 2.55 $&$  $\\
				\hline
				{$^{12}C$}&$ Reid68 $&$0.621 $&$ -2.578$&$ -8.208$&$-7.68$&$ 2.546 $&$ 2.47 $\\
				&$ UV14 $&$0.621 $&$ -1.65$&$ -7.051$&$  $&$ 2.74 $&$  $\\
				&$ AV18 $&$0.581 $&$ -0.222$&$ -5.714$&$  $&$ 2.721 $&$  $\\
				\hline
				$ ^{14}N $&$ Reid68 $&$0.621 $&$ -2.416$&$ -7.569$&$-7.475  $&$ 2.546 $&$ 2.558 $\\
				\hline	
			\end{tabular}
	\end{scriptsize}
\end{center}
\end{table*}

			
%
%
		

 \begin{table*}[ht]
	\begin{center}	
		\begin{scriptsize}
			\caption{Comparing our results with those of others' works \cite{b42,b43}.}
			
			\label{j3}
			\begin{tabular}{|l|llllll|l|}
				\hline
				$ Nuclide $&$ Potential $&$\Lambda(m_{\pi})$&$V_{E},R_{0}(fm)$& $ BE\,   (MeV)$&$ R_{rms}(fm) $&$ R_{rms}^{Exp}(fm) $&Exp.\ $(MeV)  $	\\
				\hline
				 {$^{4}He$} &$ Reid68+UIX$(This work) &$-$&$-$&$ -8.484$&$ 1.747 $&$ 1.675 $&{$-7.08$}\\
				&$ UV14+UIX$ (This work) &$-$&$-$&$ -7.761$&$ 1.747 $&$ 1.675 $&\\
				&$ AV14+UIX$ (This work) &$-$&$-$&$ -3.803$&$ 2.037 $&$ 1.675 $&\\
				&$ AV18+UIX$ (This work) &$-$&$-$&$ -5.367$&$ 1.798 $&$ 1.675 $&\\
			    &$ CD-Bonn+TM$ \cite{b34} &$4.784$&$-$&$ -7.288$&$ - $&$ - $&\\
				&$ CD-Bonn+TM$ \cite{b34} &$4.767$&$-$&$ -7.265$&$ - $&$ - $&\\
				&$ AV18+TM$ \cite{b34} &$5.156$&$-$&$ -7.210$&$ - $&$ - $&\\
				&$ AV18+TM$ \cite{b34} &$5.109$&$-$&$ -7.140$&$ - $&$ - $&\\
				&$ AV18+TM^{\prime}$ \cite{b34} &$4.756$&$-$&$ -7.09$&$ - $&$ - $&\\
				&$ Nijm\hspace{0.1cm}I+TM$ \cite{b34} &$5.035$&$-$&$ -7.150$&$ - $&$ - $&\\
				&$ Nijm\hspace{0.1cm}II+TM$ \cite{b34} &$4.975$&$-$&$ -7.135$&$ - $&$ - $&\\
				&$ AV18+Urb\hspace{0.1cm}IX$ \cite{b34} &$-$&$-$&$ -7.125$&$ - $&$ - $&\\
				&$ AV18+Urb\hspace{0.1cm}IX(GFMC)$ \cite{b35} &$-$&$-$&$ -7.075$&$ - $&$ - $&\\
				&$AV18 (HF-RPA(A)-MBPT)$ \cite{b110} &$-$&$-$&$ -8.90$&$ - $&$ - $&\\
				&$AV18 (HF-RPA(A)-MBPT)$ \cite{b110} &$-$&$-$&$ -7.90$&$ - $&$ - $&\\
				&$Reid68(CCM-FBHF3)$ \cite{b120} &$-$&$-$&$ -5.75$&$1.63$&$ - $&\\
				&$AV18(low-k)(CCM)$ \cite{b121} &$-$&$-$&$ -7.3$&$2.1$&$ - $&\\
				&$AV18(low-k)(BHF)$ \cite{b122} &$-$&$-$&$ -6.85$&$1.69$&$ - $&\\
				&$AV18(FMD)$ \cite{b123} &$-$&$-$&$ -6.99$&$1.51$&$ - $&\\
            	&{$AV18+IL7(GFMC)$ \cite{b222}} &{$-$}&{$-$}&{$ -7.105$}&{$1.43$}&{$ 1.462 $}&\\
				\hline
				{$^{6}Li$}&$ Reid68+UIX$ (This work) &$-$&$-$&$ -3.282$&$ 2.04 $&$ 2.58 $&{$-5.33$}\\
				
				&$ UV14+UIX $ (This work) &$-$&$-$&$ 0.229$&$ 2.65 $&$ 2.58 $&\\
				&{$AV18+IL7(GFMC)$ \cite{b222}} &{$-$}&{$-$}&{$ -5.3$}&{$2.39$}&{$ 2.45 $}&\\
				&$ N^{2}LO(AFDMC)$ \cite{b36} &$-$&$E\tau,1.0$&$ -5.25$&$ 2.33 $&$ 2.58 $&\\
				&$ N^{2}LO(AFDMC)$ \cite{b36} &$-$&$E{1},1.0$&$ -5.117$&$ 2.33 $&$ 2.58 $&\\
				&$ N^{2}LO(AFDMC)$ \cite{b36} &$-$&$E\tau,1.2$&$ -5.383$&$ 2.24 $&$ 2.58 $&\\
				\hline
				 {$^{12}C$}&$ Reid68+UIX $ (This work) &$-$&$-$&$ -8.208$&$ 2.546 $&$ 2.47 $&{$-7.68$}\\
			    &$ UV14+UIX $ (This work) &$-$&$-$&$ -7.051$&$ 2.74 $&$ 2.47 $&\\
			    &$ AV18+UIX $ (This work) &$-$&$-$&$ -5.714$&$ 2.721 $&$ 2.47 $&\\
			    &{$AV18+IL7(GFMC)$ \cite{b222}} &{$-$}&{$-$}&{$ -7.775$}&{$2.32$}&{$ 2.33 $}&\\
				&$ N^{2}LO(AFDMC)$ \cite{b36} &$-$&$E\tau,1.0$&$ -6.5$&$ 2.48 $&$ 2.47 $&\\
				\hline
					
			\end{tabular}
		\end{scriptsize}
	\end{center}
\end{table*}

 The total  ground state wave function of different  involved channels for each nuclei  can be  written in the form   $\Psi_{n,\ell}= a_s\psi_{2,0}+ a_d\psi_{2,2} +a_p\psi_{1,1} $. The  unknown coefficients   are obtained by  solving three equations. Two of these equations  come from evaluating the quadrupole moment, $Q=\int \psi^{\ast}(3z^{2}-r^{2})\psi d\textbf{r}$ and magnetic dipole moment of the nucleus which are matched   to the experimental values  \cite{b44}. The third one is  the normalization  of wave function, $ a_s^2 + a_d^2+a_p^2=1$.  For example,the values of $ a_s^2$, $ a_d^2$  and $ a_p^2$  for nuclide $^{6}Li$  are  $a_s^2\simeq0.934$, $a_p^2\simeq0.065$ and $a_d^2\simeq0.001$.
 \section{ \large{The three-body cluste energy } }	
\label{E3}
In this section, we calculate the three-body cluster energy using the correlation function $f_{\alpha}(r)$ and the effective potential $v_{\alpha}(r)$, which we can calculate as follows \cite{b17}:
	\begin{eqnarray}
	E_{3}&=& \frac{1}{2N}\sum_{ijk}\langle ijk|\frac{\hbar^{2}}{4m}f^{2}_{\alpha}(r_{13})\triangledown_{2}f^{2}_{\alpha}(r_{12}). \triangledown_{2}f^{2}_{\alpha}(r_{23})|ijk\rangle_{a}\nonumber\\&& \,\,\,+\frac{1}{N}\sum_{ijk}\langle ijk|h(r_{13})v_{\alpha}(r_{12})|kij-ikj\rangle\nonumber\\&& \,\,\,+\frac{1}{2N}\sum_{ijk}\langle ijk|h(r_{13})v_{\alpha}(r_{12})h(r_{23})|ijk\rangle_{a}\nonumber\\&& \,\,\,+\frac{1}{4N}\sum_{ijkl}\langle ik|h(r_{13})|jl\rangle_{a}\langle jl|v_{\alpha}(r_{12})|ik\rangle_{a},
	\end{eqnarray}
	where  $h(r_{ij})$ is as follows:
	\begin{eqnarray}
	h(r_{ij})=f^{2}(r_{ij})-1,
	\end{eqnarray}
	and $\langle ijk|\mathcal{O}|ijk\rangle_{a}$ are the antisymmetric three-body matrix element according to the harmonic oscillator wave functions.
	The results of our calculations for the mentioned nuclei using the $ Reid68 $, $AV_{14}  $, $UV_{14}$, and $AV_{18} $ potentials are given in Table \ref{j6}.
\begin{table*}[ht]
	\begin{center}	
		\begin{scriptsize}
			\caption{The three-body cluster energy.}
			\label{j6}
			
			\begin{tabular}{|c|cccc|cc|ccc|c|}
				\hline
				$Nuclide$&$$&$^{4}He$&$$&$$&$^{6}Li $&$ $&$ $&$ ^{12}C$&$$&$^{14}N $  \\
				\hline
				$Potential $&$Reid68$&$UV14$&$AV14$&$AV18$&$Reid68 $&$ UV14$&$Reid68 $&$ UV14$&$AV18$&$Reid68 $  \\
				\hline
				$\frac{E_{3}}{N}(MeV) $&$0.079$&$-0.139$&$-0.999$&$0.86$&$-0.473 $&$ -0.58$&$-0.01 $&$ -0.224$&$2.398$&$0.032 $  \\
				\hline		
			\end{tabular}
		\end{scriptsize}
	\end{center}
\end{table*}
As can be seen from the results of Table \ref{j6}, the contribution of the three-body cluster energy is about  a few KeV which is  relatively small.
	In our previous work we have shown that the three-body cluster energy has a small contribution in nuclear matter  calculations \cite{b17}. Here the same behavior  for finite nuclei can be seen with smaller contribution respect to that of nuclear matter. 

	\section{\large{Summary and conclusion}}
\label{summary}
In this work, we used a lowest order constrained variational  (LOCV) approach to obtain the ground state binding energy of some light nuclei by inclusion of the three nucleon interaction potential. In this systematic procedure, in the first step, we employed the variational method and cluster expansion to perform the calculations by using a trial wave function $ \Psi=F\Phi $, which is the product of the wave function without interaction ($ \Phi $) and the correlation function ($ F $). The correlation function  was obtained by minimizing the energy which is a functional of $F$. We used the harmonic oscillator wave function as well as the local density approximation to obtain the two-body cluster energy values for the nuclei $^{4}He$, $^{6}Li$, $^{12}C$ and $^{14}N$ with the $ Reid68 $, $AV_{14}  $, and $UV_{14}$ nuclear potentials. In the second step, we included the effect of  three-body nuclear potential in  the ground state binding energy of the above nuclei using the three-body distribution function, $g(\mathbf{r}_{1},\mathbf{r}_{2},\mathbf{r}_{3})$.
  The values of ground state  binding energy in two cases without and with the effect of three-body potential were compared for the nuclei $^{4}He$, $^{6}Li$, $^{12}C$ and $^{14}N$.
  We  also obtained the root-mean-square radius ($ R_{rms} $) of these nuclei and compared them with the experimental values.
The unknown coefficients of  wave function, $\Psi_{n,\ell}= a_s\psi_{2,0}+ a_d\psi_{2,2} +a_p\psi_{1,1} $, were also obtained by  matching the theoretical  quadrupole  and  dipole moment to those of experiment.
{A summary of conclusions according to our calculations is as follows:
}

\begin{itemize}
\item
{According to the calculations performed and the results obtained for the nuclide $^{4}He$, the amount of ground state binding energy with the $ Reid68 $ potential without the effect of three-body potential is equal to $ -4.173 MeV $ and the addition of the effect of three-body potential is $ -8.484 MeV $. The experimental value of the ground state binding energy for the nuclide $^{4}He$ is $ -7.08 MeV $ \cite{b42}, so for the $ Reid68 $ potential, we see that the amount of ground-state binding energy is closer to experiment by applying the effect of the three-body potential. Clearly, after applying the three-body potential effect, it is  seen that the energy of the ground state in the two-body potential  which is about $40\%$ off from the experimental value gets  improved to be  $16\% $ off from the experiment. We  also performed calculations with the  $ AV_{14} $, $ UV_{14} $, and $ AV_{18} $ potentials, and saw that the results obtained with these potentials are interesting. For the nuclide $^{4}He$ with the $ UV_{14} $ potential, the amounts of ground state binding energy without and with the effect of three-body interaction are equal to $ -3.603 MeV $ and $ -7.761 MeV $, respectively. Considering two-body potential, the ground state energy difference  between theoretical and experimental value   is about $49\%$. This difference reaches about $9\%$ when the effect of the three-body potential is  added. With this potential, we also see that the result is close to the experiment with the addition of the effect of the three-body potential. For the nuclide $^{4}He$ with the $ AV_{14} $ potential, the ground state binding energy without and with the three-body potential effect are $ -0.881 MeV  $ and $ -3.803 MeV $, respectively. For the $AV_{18}$ potential, the obtained values of energy without and with three-body force are equal to $-2.65 MeV$ and $-5.367 MeV$, respectively. The root-mean-square radius ($ R_{rms} $) for the nuclide $^{4}He$ with the $ Reid68 $, $ UV_{14} $, $ AV_{14} $,and $ AV_{18} $ potentials is also calculated which are $ 1.747 fm $, $ 1.747 fm $, $2.037 fm$, and $ 1.798 fm $, respectively. The experimental value of $ R_{rms} $ is $ 1.675 fm $ \cite{b43}, which is close to the values obtained for the $ Reid68 $ and $ UV_{14} $ potentials, while the value with the $ AV_{14} $ and $ AV_{18} $ potentials are not in good agreement with  the experimental value.
}

\item
{For the nuclide $^{6}Li$, as shown in the results, the amount of ground state binding energy with the $ Reid68 $ potential in the cases of without and with the three-body potential are $-1.902 MeV$ and $-3.282 MeV$, respectively.  Here, we conclude that there is a good shift in the result of calculation of binding energy to be near the experimental value, $-5.33 MeV$ \cite{b42} by adding the effect of three-body potential.
With the $ UV_{14} $ potential, the value of the ground state binding energy without the effect of the three-body potential is $ 1.5 MeV $, and by adding the effect of the three-body potential, this value is $ 0.229 MeV $. This shows that although the value with three-body potential is better than the value without three-body potential, but it is not in agreement with that of experiment. For this nuclide, the values of  $ R_{rms} $ with the $ Reid68 $ and $ UV_{14} $ potentials are $ 2.04 fm $ and $ 2.65 fm $, respectively.
We see that the value with the $ UV_{14} $ potential is close to the experimental value, $ 2.58 fm $ \cite{b43}.
}

\item
 {For the nuclide $^{12}C$, according to the results, the amount of ground state binding energy with the $ Reid68 $, $ UV_{14} $,and $ AV_{18} $ potentials without the three-body potential are $ -2.578 MeV $, $ -1.65 MeV $ and $-0.222 MeV$, and with the three-body potential are equal to $ -8.208 MeV $, $ -7.051 MeV $,and $-5.714 MeV$. 
	 Therefore, the three-body potential is very effective in the amount of binding energy to nearly agree with experimental value for this nuclide, ($ -7.68 MeV $ \cite{b42}).  The values of the $ R_{rms} $ with the $ Reid68 $, $ UV_{14} $, and $ AV_{18} $ potentials are equal to $ 2.546 fm $, $ 2.74 fm $,and $2.721 fm$, respectively, which are close to the experimental value, $ 2.47 fm $ \cite{b43}.
}

\item
{For the nuclide $^{14}N$,
according to the results, we saw that the inclusion of three-body potential leads to a substantial change for the ground state binding energy
 from $-2.416 MeV$ to $-7.569 MeV$ which  is close to the experimental value, $ -7.475 MeV $ \cite{b42}. By using the three-body potential effect, we saw a $70\%$ improvement in the ground state binding energy of the $^{14}N$ nuclide.
The value of the $ R_{rms} $ with the $ Reid68 $ potential is equal to $ 2.546 fm $, which are close to the experimental value $ 2.558 fm $ \cite{b43}.
}

\end{itemize}

Finally, as discussed above, it was seen that the three-body potential generally has improved the results of our calculations. This means that by adding its effect in the calculations, the ground state binding energies for the nuclei $^{4}He$, $^{6}Li$, $^{12}C$ and $^{14}N$ have an overall agreement with those of experiment. Therefore, the three-body interaction can effectively improve the results for ground state binding energy.
It should be also mentioned that our results were also compared with those of other works. 
%
{Here a notable point is that we were able to obtain reasonable results by performing simple calculations compared to unconstrained methods such as FHNC method that parametrize the short-range behavior of correlation function attempting to go beyond the lowest order \cite{b201}.}


\section*{Acknowledgments}
We wish to thank Shiraz University Research Council.


\end{document}